\documentclass{iopart}
\usepackage{graphicx}
\usepackage{color}
\usepackage{amssymb}
\usepackage{fancyhdr}
\begin{document}

\title{Growth exponents of the etching model in high dimensions}
\author{Evandro A Rodrigues, Bernardo A Mello, Fernando A Oliveira }
\address{Institute of Physics, University of Brasilia, Caixa Postal 04513, 70919-970 Brasilia-DF, Brazil}
\ead{evandro@fis.unb.br}

%\author{Evandro A. Rodrigues }
%\author{Bernardo A. Mello }
%\author{Fernando A. Oliveira}
%\email{faooliveira@gmail.com}
%\affiliation{Institute Of Physics, University of Brasilia, Caixa Postal 04513, 70919-970 Brasilia-DF, Brazil}

\begin{abstract}

In this work we generalise the  etching model~\cite{Mello2001} to $d+1$ dimensions.
The dynamic exponents of this model are compatible with those of the KPZ Universality Class.
We investigate the roughness dynamics  with surfaces up to $d=6$ . We show that
the data from all substrate lengths and for all dimension can be collapsed in to one
common curve. We determine the dynamic exponents as a function of the dimension.
Moreover, our results suggest that $d=4$ is  not an upper critical dimension for the etching model, and that
it fulfills the Galilean Invariance.\\

Published as: J. Phys. A: Math. Theor. 48 (2015) 035001

doi:10.1088/1751-8113/48/3/035001
\end{abstract}

\submitto{\JPA}

%\date{\today}
\maketitle
%\ioptwocol
\section{Introduction}

Over recent decades, surface dynamics has become an intense research topic in
complexity. This interest has partially been motivated by the description of several real
world system dynamics such as stochastic processes,  weather and population dynamics. Under the statistical approach,
the detailed structure of these systems is abstracted, and  modelled  in
simple ways.  Those models  have been grouped into universality classes of similar dynamics
\cite{Family1985}.

Of particular interest in the study of growing surfaces is the KPZ
equation~\cite{Kardar1986}
\begin{equation}
\partial_{t} h(\vec{x},t) = \mu \nabla^2 h + \frac{\lambda}{ 2} (\vec{\nabla} h)^2 + \eta(\vec{x},t) .
\end{equation}
It describes the kinetics of a $d$ dimensional surface in a space of $d+1$
dimensions. The roughness smoothing is controlled by the surface tension $\mu$,
$\lambda$ quantifies the non-linear growth and $\eta$ is the non-correlated
noise. Besides its theoretical applications, the KPZ equation describes many real
world systems, such as flame front propagation \cite{Provatas1995} and
deposition of thin films \cite{Lita2000}.

   For the particular case of the KPZ in $1+1$ dimensions, the dynamic exponents
have been known for more than twenty years~\cite{Hwa1991,Frey1996}
 and one exact solution was recently discussed by Sasamoto et al~\cite{Sasamoto2010}. 
However, there are no such exact results for higher dimensions. 
In these cases the growth exponents are obtained by two approaches:
 numerical simulations and approximate methods.

The non-linear character of the KPZ equations leads to behaviors similar to those
found in cellular automata models of atomistic surfaces. Some of the recent
developments include extensions on the Eden Model \cite{Hosseinabadi2010}, use of
the high parallelism of GPU computation \cite{Kelling2011} and numerical
integration of the KPZ equation \cite{Miranda2008}. Since simulations of large
substrates and higher dimensionality are memory bounded and very time consuming,
numerical approaches are limited by hardware availability.

By use of renormalization group techniques it has been conjectured that the KPZ 
equation is the field theory of many surface growth models, such as the Eden model,
ballistic deposition, the RSOS model and the PNG model\cite{Sasamoto2004}. A rigorous proof has been provided by Bertini
  and Giacomin \cite{Bertini1997} in the case of the RSOS model.

Most stochastic models of surface dynamics with relaxation present the behavior
described by the Family-Vicsek (FV) relation~\cite{Family}. As shown in figure
\ref{fig:dynamics15d}, the roughness $w(t)$ grows initially as a power-law
function of time, with exponent $\beta$. When $t \rightarrow \infty$ the roughness
saturates at 
\begin{equation}\label{eq:lalpha}
w_s \propto L^\alpha,
\end{equation}
 where $L$ is the substrate size. These
properties are expressed in the FV relation:
\begin{equation}\label{eq:FV}
%amsmath
%  w(t,L) = w_s f(t/t_\times,\beta)= \begin{cases}
%    w_y t^\beta  & \text{if $t \ll t_\times$}\\
%     w_s & \text{if $t \gg t_\times$}\\
%  \end{cases}.
  w(t,L) = w_s f(t/t_\times,\beta)= \left\{ \begin{array}{ll}
     w_y t^\beta  & \mbox{if  $t \ll t_\times$}\\
     w_s & \mbox{if  $t \gg t_\times$} 
  \end{array} \right.
\end{equation}
The two regimes are separated by the saturation time $t_\times$, defined as the
intersection of the two functions above, leading to  
\begin{equation}\label{eq:lz}
t_\times \propto L^z,
\end{equation}
with $z=\alpha/\beta$.

Bearing in mind the attention obtained by the KPZ  and related model, 
even if only numerical solutions were considered,
one would expect most of its dynamics should be fully understood by now. 
However, that is not the case. For example,
the absence of exact solutions leads to the much debated possibility of an
upper critical dimension (UCD) $d_c$ for the dynamic exponents.
 For a concise review see  \cite{Schwartz2012}.

From the results of numerical experiments, mathematical expressions involving integers
have been proposed to describe how the dynamic exponents of models belonging 
to the KPZ universality class depend on the dimension $d$, i.e. $\alpha=\alpha(d)$, 
$\beta=\beta(d)$ and $z=z(d)$. The best known are those for the RSOS model,  by Kim and 
Kosterlitz~\cite{Kim1991}, for the Eden model \cite{Moser1991} by  Kerst\'esz and Wolf,  
the heuristic approach to the strong-coupling regime by Stepanow \cite{Stepanow1997}
and a tentative method based on  quantization of the exponents by  L\"assig\cite{Lassig1998}.
Unfortunately, further numerical results have shown that these expressions are 
neither exact nor precise \cite{Forrest1990,Tang1992,Ala-Nissila1993,Ala-Nissila1998}.

 Analytical methods such as mapping of the directed polymer \cite{Halpin-Healy1989}, 
perturbation expansion \cite{Wiese1998} and mode-coupling techniques
\cite{Bhattacharjee1998} among others
\cite{Lassig1995,Lassig1997,T.Blum1995,M.A.MooreT.BlumJ.P.Doherty1995} suggest the value
$d_c = 4$.
 On the other hand, such a limit was not found by numerical studies
\cite{Moser1991,Marinari2000,Marinari2002}, or by the numerical and theoretical results 
obtained by Scharwartz and Perlsman \cite{Schwartz2012}.

In this work we contribute to the discussion regarding dynamic exponent values
by extending the Etching Algorithm by Mello et al \cite{Mello2001} 
to $d+1$ spatial dimensions. 
The exponents obtained in those works by simulations of this model are mostly 
compatible with the  values of the KPZ equation.

Here we determine the exponents for $1\le d\le 6$ and we compare them with other 
numerical results in the existing literature, concluding that  if the UCD exists, 
it is no less than $d_c = 6$. Moreover, we show that this version of the etching
model obeys the Galilean Invariance.

\section{The etching model in $d$ dimensions}

Surface roughness obtained by numerical simulation of a discrete atomistic model
often presents the FV scaling. Examples are the ballistic deposition (BD)
\cite{vold1963computer} and the Wolf-Villain \cite{Wolf1990} model (WV).

One of those models is the etching algorithm, which is a simple atomistic
 model that mimics the etching of a crystalline solid by a liquid~\cite{Mello2001}. It
was originally proposed for $d=1$, in which case the scaling exponents are
very close to those of the KPZ equation, namely $\alpha = 0.4961 \pm 0.0003$
and $\beta = 0.330\pm0.001$. For this reason the model is believed to belong
to the KPZ universality class, although this has not been formally proven.

Some properties of this model have been investigated,
such as its Kramers-Moyal coefficients and Markov length scale\cite{Kimiagar2008},
maximum and minimum height distribution\cite{Oliveira2008}, height and 
roughness distributions in thin films\cite{Paiva2007} and a variation of the model
on $2+1$ dimensions used to test a novel method for roughness exponent 
estimation\cite{AaraoReis2006a}.

In the present work we investigate the dynamics
and the exponents of the roughness of this model extended to $d+1$ dimensions.
%Although dynamic exponents of the KPZ universality class in $1+1$ dimensions are analytically known, these values are still to be found for higher dimensions.
For this model, the ``solid'' is a square lattice exposed to a solvent, and the
removal probability of each cell is proportional to its exposed area. The
cellular automata with $d=1$ is
\begin{enumerate}
\item at discrete instant $T$ one horizontal site $i=1,2...,L$ is randomly
    chosen;
\item $h_i(T+1) = h_i(T)+1$;
\item if $h_{i+\delta}(T) < h_i(T)$, do $h_{i+\delta}(T+1) = h_i(T)$, where
$\delta=\pm 1$ are the first neighbours. 
\end{enumerate}
The general case $i$ and $\delta$ are vectors and $\delta$ runs over the
$2^d$ first neighbours of the hypercube. If $L$ is the substrate length in each
direction, the total number of sites is $L^d$. The normalized time $t$ defines
the time unity as $L^d$ cellular automata iterations, i.e., $t=T/L^d$.

Note that (i) and (ii) introduce randomness in time and space, 
this is equivalent to the noise $\eta$ in the KPZ equation.
The  off-diagonal condition (iii), combines  the linear Laplacian term -- 
 which tends to smooth the surface reducing its curvature -- with a nonlinear term, the lateral growth,
  equivalent to the Burgers equation. 
Of course this cellular  automata is not the KPZ, but we expect it to mimic the KPZ dynamics for $d+1$ 
dimensions as it does for $1+1$.

\begin{figure}
\center
\includegraphics[width=0.6\columnwidth]{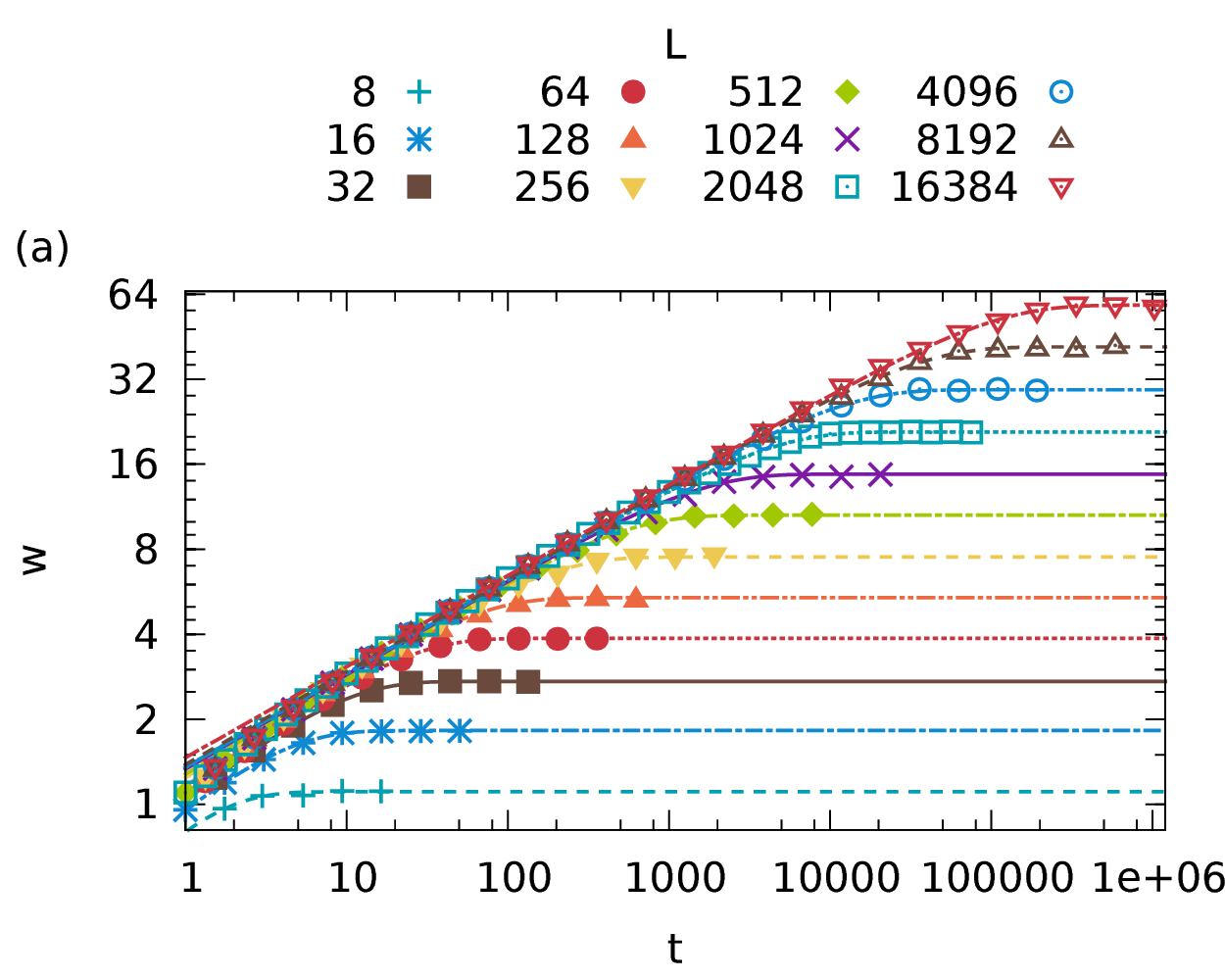}\\
\includegraphics[width=0.6\columnwidth]{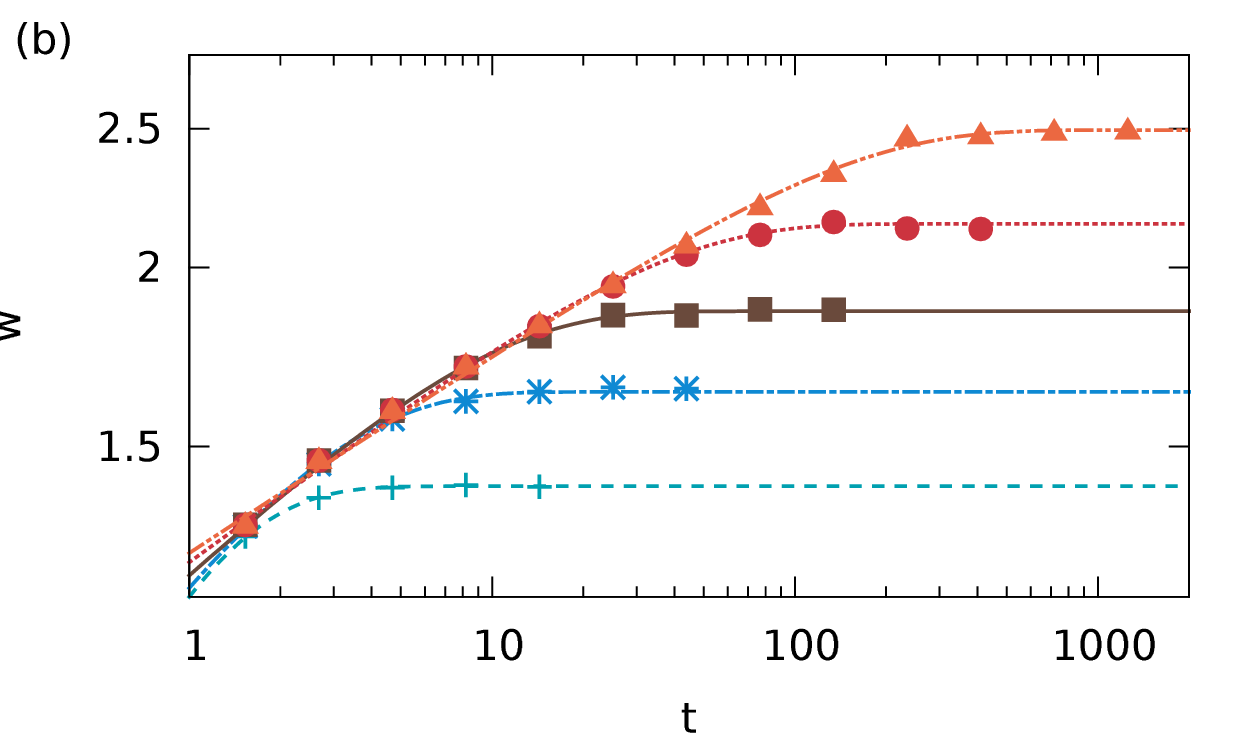}\\
\includegraphics[width=0.6\columnwidth]{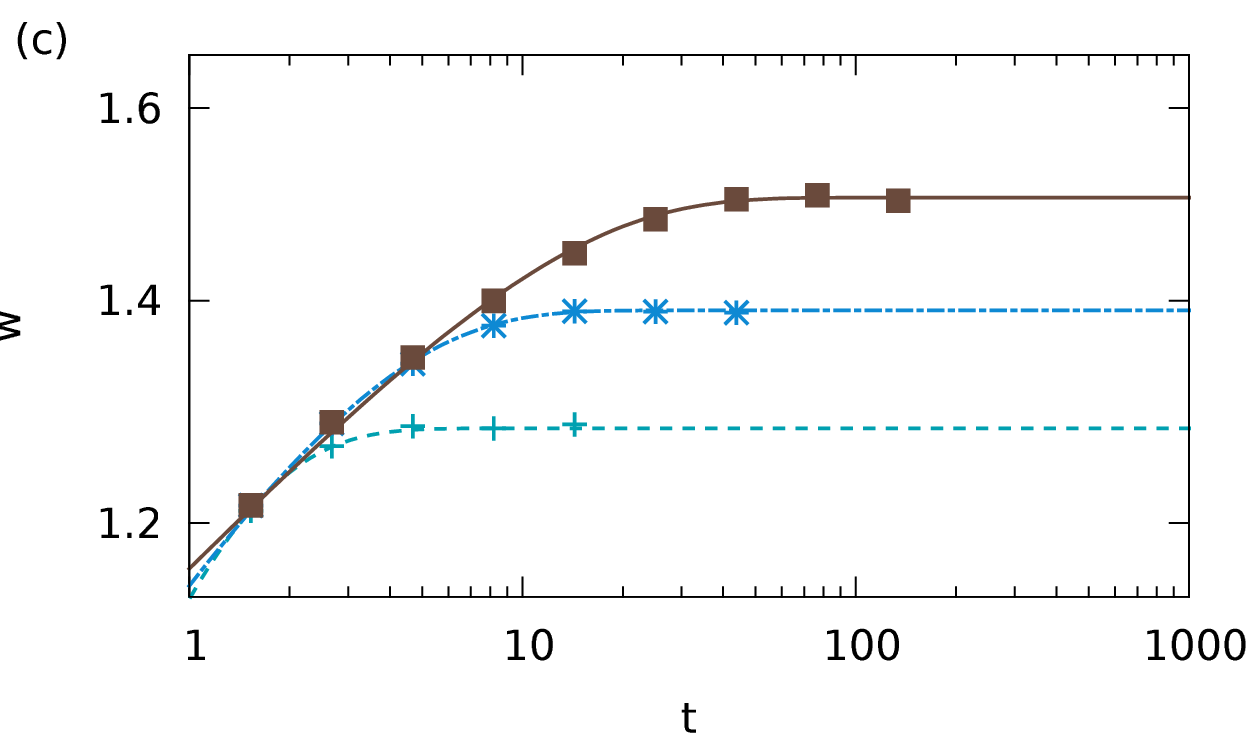}
\caption{
Roughness $w(L,t)$ from the etching model as a function of time, for surfaces with
(a) $d=1$, (b) $d=4$, and(c) $d=6$, showing only data points for $t < 15 t_\times$ for
clarity. The lines are guides for illustration.
}
\label{fig:dynamics15d}
\end{figure}

For each dimension $d$, the simulations are performed with several substrate lengths $L$. For each value of $d$ and $L$, the experiment is repeated several
times, and the ensemble average taken to reduce noise. As commonly done in surface
dynamics simulations, we apply periodic boundary conditions to reduce the unwanted
finite length effects.

Figure \ref{fig:dynamics15d} shows the roughness evolution for some
substrate lengths of dimensions $d=1$, 4 and 6 in a log-log scale. The
same behavior appears for $d=2$, 3 and 5 dimensions. The short
range correlations develop at very small time scales $t\lesssim 1$, resulting
in deviations from the FV.   This transient plays an important role in determining
the growing exponents, mainly for higher dimensions, where computer power limitations
 impose small values of $L$, leading to small values of $t_\times$.
After that time all substrates show
the expected power law like behavior for $t \ll t_\times$, saturating when
$t\gg t_\times$.

\section{The data collapse}

Later in this work we obtain the parameters $w_s$, $\beta$ and $t_\times$  from our simulation data.
Once the values of $w_s$ and $t_\times$ are known for every $L$ of
a given $d$, the corresponding data may be rescaled,
 resulting in  the collapse predicted by
the FV relation (\ref{eq:FV}).

\begin{figure}
\center
\includegraphics[width=0.70\columnwidth]{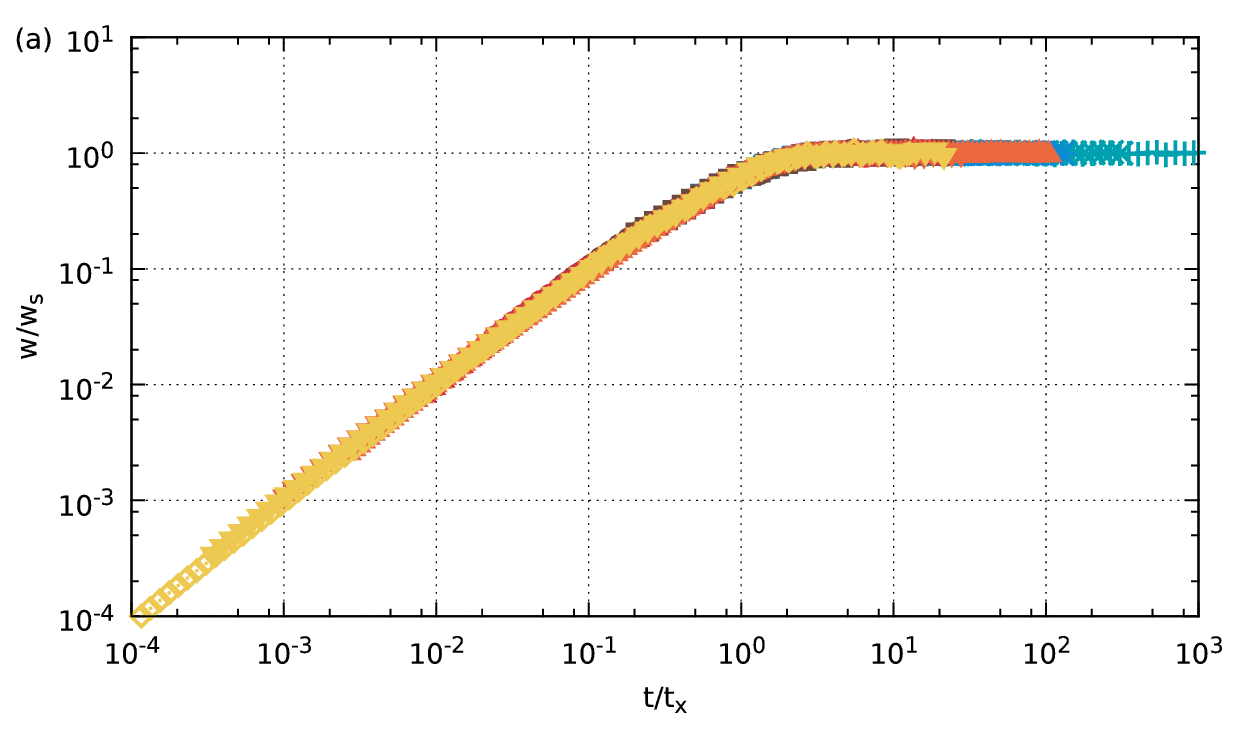}\\
\includegraphics[width=0.70\columnwidth]{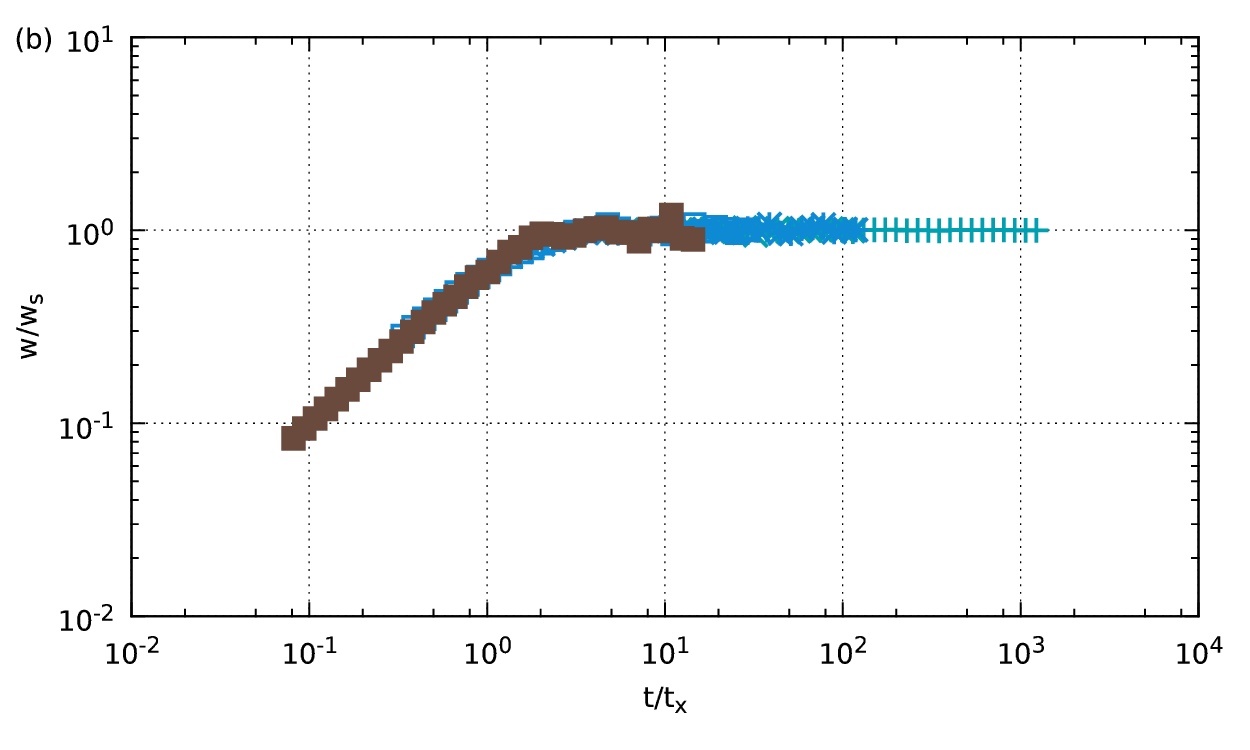}\\
\includegraphics[width=0.70\columnwidth]{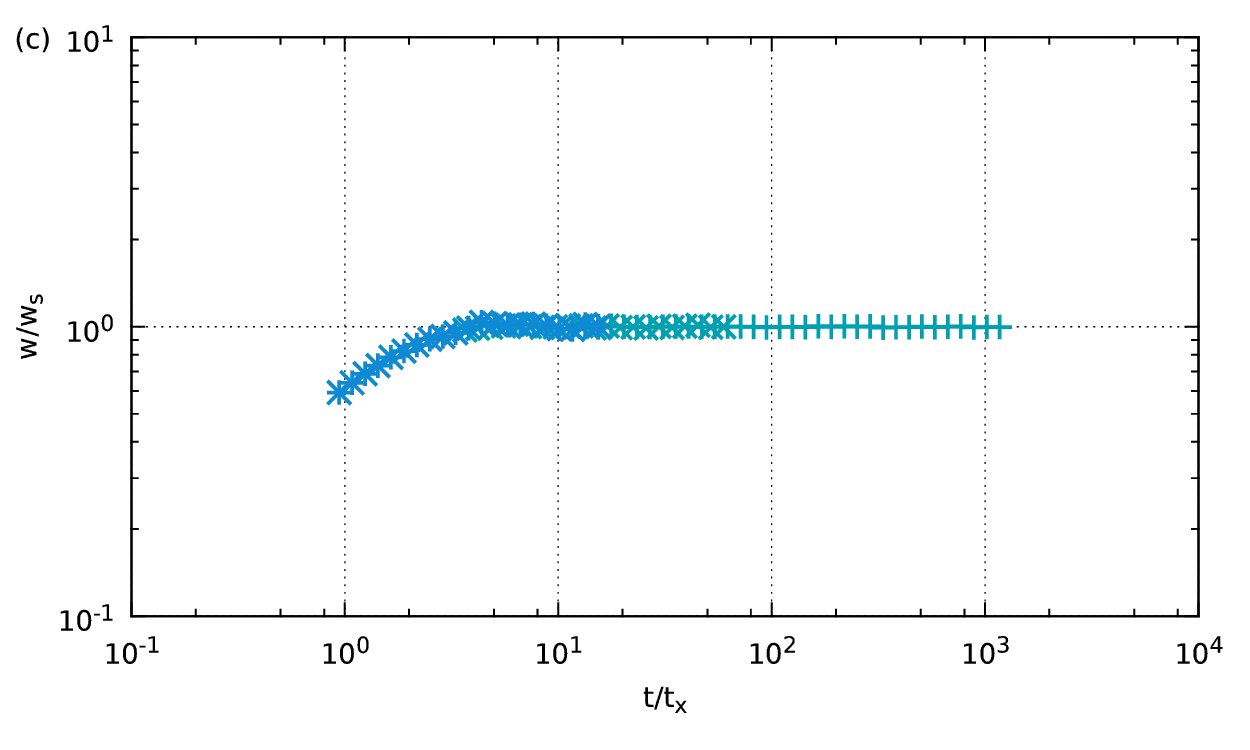}
\caption{Collapse of the data from all values of $L$
for each dimension $d$, (a) $d=1$, (b) $d=4$, and(c) $d=6$. The collapse was achieved
by applying the scales indicated in the labels of the axis.
Only points with $t>10$ were included in the above plot to exclude
the transients at $t\lesssim 1$.}
\label{fig:dynamicsCollapse}
\end{figure}

The parameters $\beta$, and $t_\times$ are responsible for the collapse
at $t\ll t_\times$, while the parameter $w_s$ is responsible for the collapse at $t\gg
t_\times$. If these parameters are to be obtained from fitting the roughness
as a function of the time for each value of $d$ and $L$, then this is the minimal
set of parameters required to collapse the data. The agreement is verified in both
extreme regions of figures \ref{fig:dynamicsCollapse}a and
\ref{fig:dynamicsCollapse}b, indicating that these parameters were properly
obtained.

\section{Finding the dynamic exponents}

Power law fitting (PL) is a common method used to determine the parameters of
 (\ref{eq:FV}). It consists of fitting the values of $w_y$, $\beta_L$, and
$w_s$ by using the two expressions of  (\ref{eq:FV}) at $t\ll t_\times$ and
$t\gg t_\times$. After this, the value of $t_\times$ can be determined by
the intersections of the functions of the two regimes. 

In this work, we refine the usual power law fitting by using a numerical iterative 
data collapse. Our method consists of initially obtaining a set of exponents and using
this initial set as a seed for the next curve. Using this simple technique, we observe
a significant increase in precision for the case where $d=1$, for which exact exponents 
are known.

The transient at $t\lesssim 1$ is a problem for small substrates, for which the value of
$t_\times$ may be so low that the roughness saturation mixes with the short range correlation,
 affecting the evaluation of $\beta$ and $t_\times$. For that reason, we do not expect the 
 parameter $\beta$  to be independent of $L$. To make this clear, we use the
subscript $L$ in the parameter $\beta_L$ obtained in such a way.

%\section{Finding the dynamic exponents}

\begin{figure}
\center
\includegraphics[width=0.7\columnwidth]{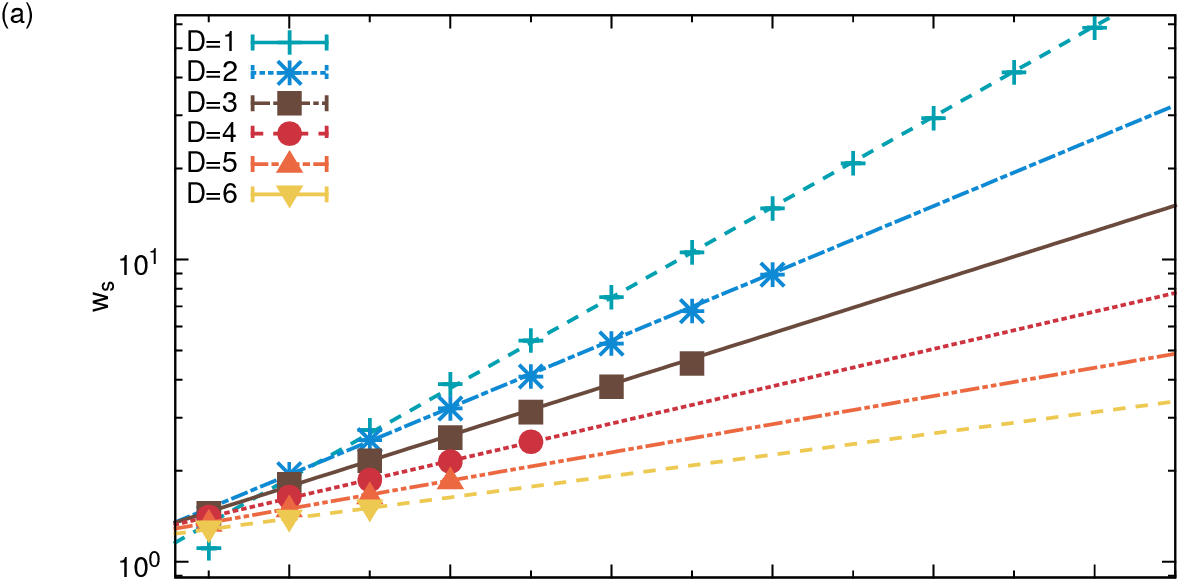}\\
\includegraphics[width=0.7\columnwidth]{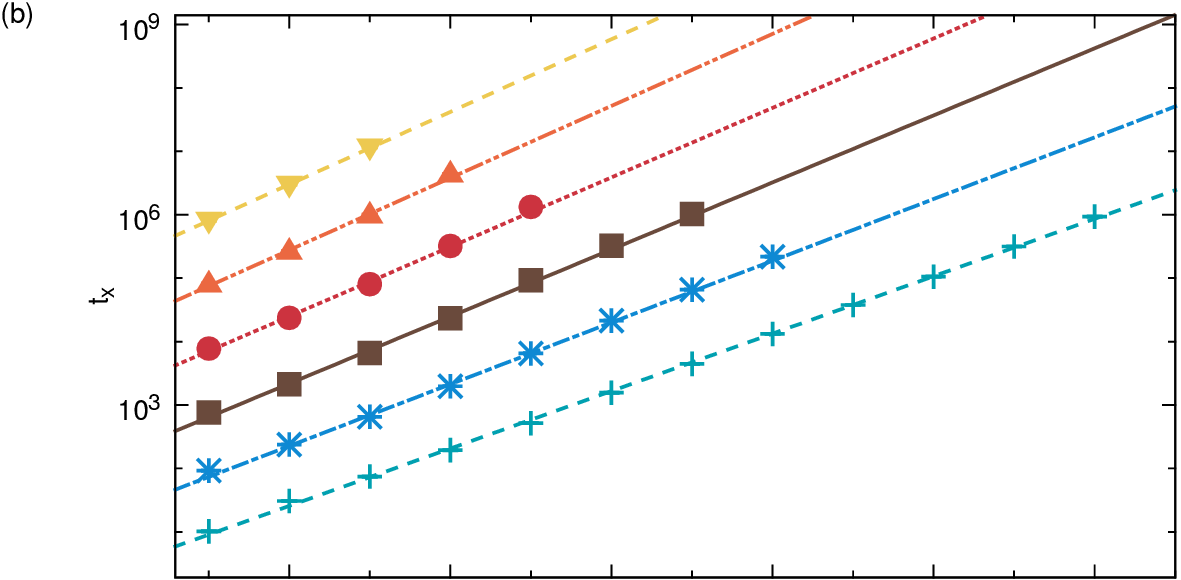}\\
\includegraphics[width=0.7\columnwidth]{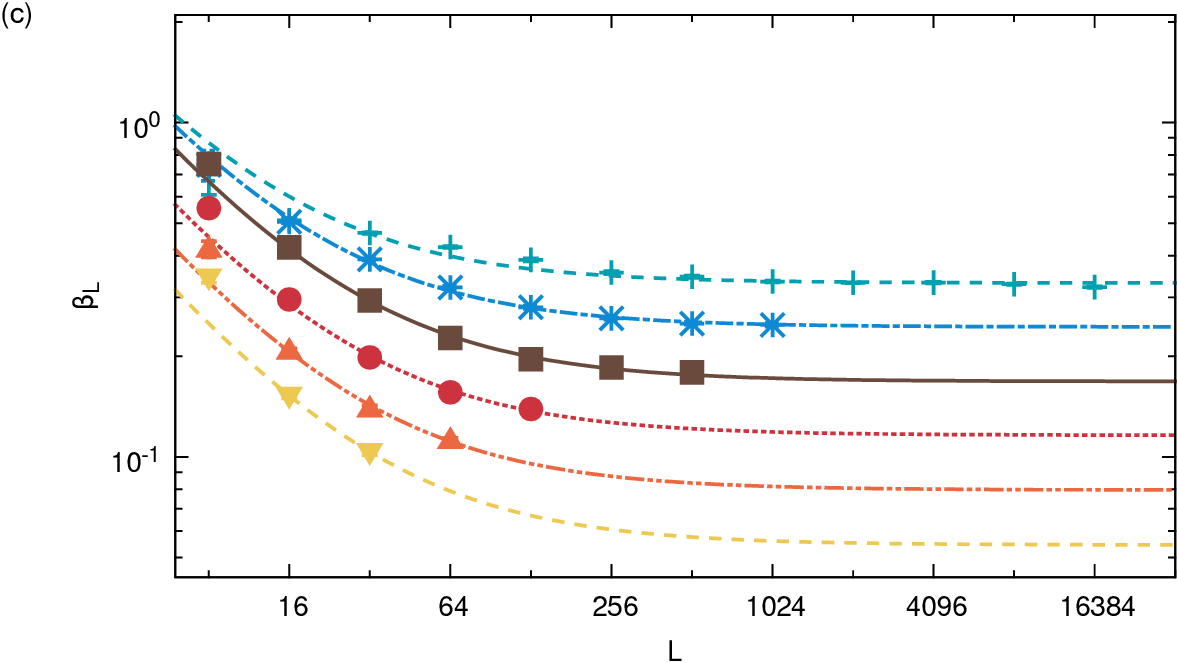}
\caption{Parameters of (\ref{eq:FV}) plotted as functions of $L$ with $d=1\dots6$.
(a) $w_s(L)$, (b) $t_\times(L)$ multiplied by $10^d$ for better visualization,
and (c) $\beta_L(L)$. In all simulations $L=2^n$, with $n$ integer.}
\label{fig:dim_ws_tx_beta}
\end{figure}

The values of $\beta_L$, $w_s$ and $t_\times$ for each value of $d$ and $L$
were obtained from roughness fitting and plotted in figure
\ref{fig:dim_ws_tx_beta}.

As prescribed by the FV, $w_s \propto L^\alpha$ and $t_\times \propto L^z$, for
a big enough $L$. When substrates are small, $t_\times \approx 1$,  the power laws are 
disturbed by the transient behavior occurring  at this time scale.
Therefore, the exponents must rely more strongly on the points with  higher values of $L$.

When performing the fitting of figure \ref{fig:dim_ws_tx_beta} we incorporated the
error obtained from the roughness fitting. The fitting error is a combination
of the error due to the stochastic data fluctuation and the error due to the
deviation between the fitting function and the data. 
Since that deviation is strong at
$t\lesssim 1$, the use of the fitting error from the roughness data in the fittings of figure
\ref{fig:dim_ws_tx_beta} reduces the weight of the points witha  small $L$,
because, in these cases, a significant part of the data has small values of $t$.

The fitting of power laws to the points of figures \ref{fig:dim_ws_tx_beta}a-b
leads, respectively, to the exponents $\alpha$ and $z$.

Because $\beta_L$ is the main parameter controlling the curve shape at small values of $t$,
 it is also the parameter most strongly disturbed by the initial transient.
 If we define $\beta$ as the asymptotic value of $\beta_L$,
it can be found by using the points from figure~\ref{fig:dim_ws_tx_beta}c to do
the finite size scaling
\begin{equation}
\beta_L = \beta \left(1+\frac{A_0}{L^\gamma}\right). \label{beta}
\end{equation}
%where $\gamma \approx 1$.From now on we will use the size independent value $\beta$. 
Where $\gamma \approx 1$ and $A_0$ are parameters to be adjusted. From now on we will use the size independent value $\beta$.

\section{ Universality}

\begin{figure}
\center
\includegraphics[width=0.90\columnwidth]{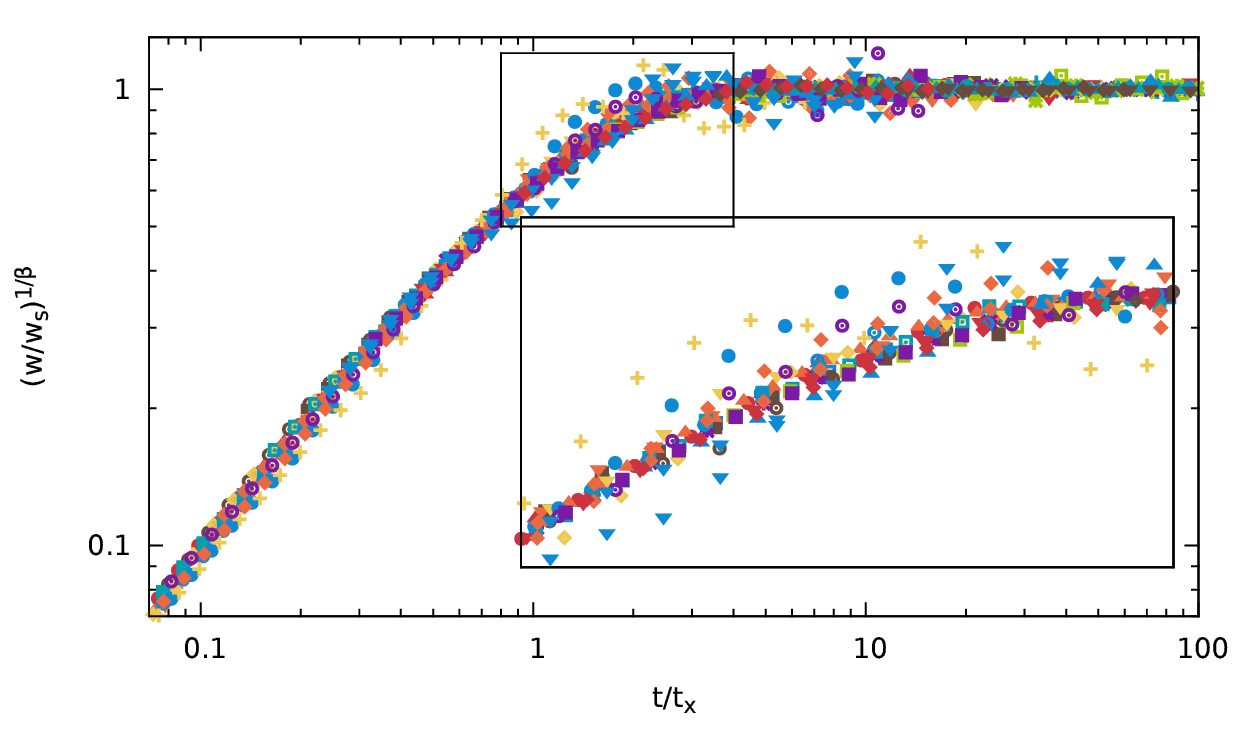}
\caption{Collapse of the data from all values of $L$
for all dimensions $d$. The collapse was achieved by applying the scales indicated in the labels of the axis.
The inset contains a zoom around the saturation time $t_x$.
}
\label{fig:dynamicsCollapseAll}
\end{figure}

In the last section we have seen that for a given dimension $d$ when we scale the axes 
as $w \rightarrow w/ w_s$ and $t \rightarrow t/t_\times$ all curves collapse in a single 
one. Since we know that the exponentes depends on the dimension. We conjecture how would 
those curves behave if now we scale the axes as $( w/ w_s)^{\frac{1}{\beta}}$ and $t/t_\times$? 
The results are exhibited in figure~\ref{fig:dynamicsCollapseAll}, where we plot all 
the curves for all dimension. We use a method of minimization of the error and from 
that we can obtain the collapse of all data, i.e. for all dimensions $d$ and for all 
lengths $L$. 
That seems to be an universal behaviour.
%We see that we get an universal behavior. 
Moreover, the exponents obtained 
by this method are much more precise than when we have the collapse for just on single dimension $d$.
It is notable that even considering that the $y$ scale should amplify statistical noise, the data
from all curves collapses very well around the saturation time $t_\times$, as well as the extremes
for $t/t_\times \rightarrow \infty$ and $t/t_\times \rightarrow 0$

These results suggest that $( w/ w_s)^{\frac{1}{\beta}}=g(t/t_\times)$ where the function $g(x)$ satisfy
\begin{equation}\label{eq:FV2}
%  w(t,L) = w_s g(t/t_\times)^{\beta}= \begin{cases}
%    w_y t^\beta  & \text{if $t \ll t_\times$}\\
%     w_s & \text{if $t \gg t_\times$}\\
%  \end{cases},
  w(t,L) = w_s g(t/t_\times)^{\beta}= \left\{ \begin{array}{ll}
     w_y t^\beta  & \mbox{if $t \ll t_\times$}\\
     w_s & \mbox{if $t \gg t_\times$}\\
  \end{array}\right.,
\end{equation}
i.e. the FV relation.

\section{ Upper critical dimension and Galilean Invariance}

Before presenting our results we present in Tables \ref{table:wk_kk_me_compare} and 
\ref{table:wk_kk_me_compare_beta} the values of $\alpha$ and $\beta$ from other authors.
Ala-Nissila\cite{Ala-Nissila1998} performed simulations of the RSOS growth model for dimensions $d \geq 4$.
Marinari et al \cite{Marinari2000} use a RSOS discretization of the surface, finding values for surfaces of up to $4+1$ dimensions.
Katzav And Schartz\cite{Katzav2004} obtained these exponents for ballistic deposition. Pagnani and Parisi\cite{Pagnani2013} use 
multisurface coding on four dimensional RSOS model.
\'{Odor} et al\cite{Odor2010} map growth models onto driven lattice gases of $d$-mers.
Canet et al \cite{Canet2010} developed a simple approximation of the non-pertubative renormalization group for the KPZ
equation.
Tang et al \cite{Tang1992} propose a hypercube-stacking model.

\begin{table}[!ht]
\caption{ Growth exponent $\alpha$ as obtained from several authors. \label{table:wk_kk_me_compare} }
\centering
\begin{tabular}{ c c c c c c c }
\mr
$d$  & Ala\cite{Ala-Nissila1998}   & Ma\cite{Marinari2000}   & Ka\cite{Katzav2004}    & Od\cite{Odor2010} & Ca\cite{Canet2010}  & Pa\cite{Pagnani2013}        \\
\br
$1$  & $ -   $               & $ -        $            & $ 0.45 $                     & $ -      $        & $ 0.50 $      	&    		\\
$2$  & $ -   $               & $0.393(3)  $            & $ 0.26 $                     & $0.395(5)$        & $ 0.33 $      	&    		\\
$3$  & $ -   $               & $0.3135(15)$            & $ 0.12 $                     & $0.29(1) $        & $ 0.17 $      	&    		\\
$4$  & $ 0.141(1)$           & $0.255(3)  $            & $ - $                        & $0.245(5)$        & $ 0.075$      	&$0.2537(8)$    \\
$5$  & $ -   $               & $ -        $            & $ - $                        & $0.22(1) $        & $ - $         	& 		\\
$6$  & $ -   $               & $ -        $            & $ - $                        & $ -      $        & $ - $         	& 		\\
\br
\end{tabular}
\end{table}

\begin{table}[!ht]
\caption{ Growth exponent $\beta$ as obtained from several authors. \label{table:wk_kk_me_compare_beta} }
\centering
\begin{tabular}{ c c c c }
\hline
$d$  & Ta\cite{Tang1992}   & Ala\cite{Ala-Nissila1998} & Od\cite{Odor2010}  \\
\hline
$1$  & $ 0.333(1) $        & $ -        $              & $ 0.333(5) $  \\
$2$  & $ 0.240(1) $        & $ -        $              & $ 0.240(1) $  \\
$3$  & $ 0.180(5) $        & $ -        $              & $ 0.184(5) $  \\
$4$  & $ -   $             & $ 0.16(1)  $              & $ 0.15(1)  $  \\
$5$  & $ -   $             & $ 0.11(1)  $              & $ 0.115(5) $  \\
$6$  & $ -   $             & $ 0.09(1)  $              & $ - $  \\
\hline
\end{tabular}
\end{table}

Table \ref{table:dfg_results} shows the values of $\alpha$, $\beta$, and $z$
obtained as described in the previous paragraphs for the etching model. 
%%The fifth  column
%%contains the relative difference between $z$ and $\alpha/\beta$, i.e.
%%\begin{equation}
%%\Delta z \equiv 2|z-\alpha/\beta|/(z+\alpha/\beta), \label{dz}
%%\end{equation}
%%which should be $0$ if $\alpha$, $\beta$ and $z$ were exact.  
Undoubtedly, the exponents are not constant for $d>4$. Altogether, our results suggest that there is not an UCD
$d_c \le 6$, in agreement with previous results
\cite{Moser1991, Marinari2000, Marinari2002,Schwartz2012}.

The renormalization group yields a version of the Galilean Invariance (GI)~\cite{barabasi} for the KPZ equation
as 
\begin{equation}\label{eq:GI}
\alpha + z = 2. 
\end{equation}
This relation is considered to be true for all dimensions,
making it a reliable metric of exponent values for higher dimensions. The universality of this
relation has been questioned in a recent work \cite{Wio2010} where Wio et al demonstrate that it is 
possible for a system to show KPZ scaling without obeying the GI.

On the sixth column of table \ref{table:dfg_results} we present our values of $\alpha + z $.
It is difficult to compare our results concerning the GI with other authors because some of them use
$z=\alpha/\beta$ and the GI to produce $\beta$ and $z$. We have determined those exponents independently. 
The  dynamic exponents obtained here are precise enough, up to $5$ dimensions,
allowing  us to conclude that the GI holds for the etching model.
However, it is interesting to note that previous works by Tang et al \cite{Tang2010,Xun2012}
analysed the etching model on surfaces with a fractional dimension such as the Sierpinski
Carpet ($d=1.465$), the Sierpinski Arrowhead and the crab ($d=1.585$). 
On both works, the expected condition of $ \alpha + z = 2 $ is not confirmed 
for such dimensions. This is worth to mention, however, it is not a surprise since fractal geometries
are not continuos.

\begin{table}
\caption{Dynamic exponents obtained from the fittings of figure
\ref{fig:dim_ws_tx_beta}. A evidence of the precision of these exponents is the 
value of $\alpha+z$, which should be 2. 
\label{table:dfg_results}}
\centering
\begin{tabular}{ c c c c c } 
\hline
$d$           &$\alpha  $  &$\beta   $  &$z$          &$\alpha+z$ \\
\hline
\hline
$1$     & $0.497(5)$    &$0.331(3)$     &$1.50(8)$    & $2.00(1)$ \\
$2$     & $0.369(8)$    &$0.244(2)$     &$1.61(5)$    & $1.98(2)$ \\
$3$     & $0.280(7)$    &$0.168(1)$     &$1.75(9)$    & $2.03(2)$ \\
$4$     & $0.205(3)$    &$0.116(3)$     &$1.81(3)$    & $2.02(1)$ \\
$5$     & $0.154(2)$    &$0.079(3)$     &$1.88(6)$    & $2.04(1)$ \\
$6$     & $0.117(1)$    &$0.054(1)$     &$1.90(6)$    & $2.01(1)$ \\
\end{tabular}
\end{table}

\section{Conclusion}

We  have successfully  generalized  the  etching model for $d +1$ dimensions, which  has characteristics
similar to KPZ model, and also obtained the growth exponents for $d \leq 6$ and have shown the data 
 collapse for several substrate lengths and dimensions 
 see figure \ref{fig:dynamicsCollapseAll}
This was obtained through a simple scaling of the axes,  resulting in  a perfect collapse of the data, at least up to $d=6$. 

We studied the dependence of the dynamic exponents with the dimension $d$.   Our data
suggest that there is no UCD  $d_c=4$ for the etching model.  
The  exponents obtained here are precise enough to allow us to conclude that the GI holds for  dimensions $d \leq 6$ integer. 
It is interesting to mention that the etching model on surfaces 
with a fractional dimension\cite{Tang2010,Xun2012} could violate the GI.  
If the Galilean Invariance is violated on fractal dimensions, such result should not 
be a surprise, since there is no continuum space transformations in fractal geometry.
%This should not be a surprise, 
% since there is no continuous space transformation in a fractal geometry.
%This is another matter for future research.

To date, we have not presented a proof that the etching model is equivalent to the KPZ,  
such as the proof given by Bertini and Giacomin \cite{Bertini1997}  for the RSOS model. 
The combination of the renormalization of probability of height distribution in a 
lattice \cite{Oliveira2000} with recent scaling for asymptotic times \cite{Ferreira2012} could 
yield some result to this problem. Moreover we expect as well that we could obtain the function $g$ discussed in the section $V$. We are working to show that in subsequent publications. 

More than $20$ years after the KPZ  seminal work we still do not have a final solution 
for some important questions, nevertheless
very good numerical methods and theoretical approaches have been suggested, 
which clearly indicates that this is still a very rich research  field.
We also hope that this work stimulates new research toward solutions to those problems.

%\section{Acknowledgements}
\ack
This work was supported by the Conselho Nacional de Desenvolvimento Cientifico
e Tecnologico (CNPQ), the Coordena\c{c}\~ao de Aperfei\c{c}oamento de Pessoal
de N\'ivel Superior (CAPES), the Funda\c{c}\~ao de Apoio a Pesquisa do Distrito
Federal (FAPDF), and the Companhia Nacional de Abastecimento (CONAB).
%

%\bibliographystyle{unsrt}
%\bibliography{bibliography,mendeley}
\end{document}